\RequirePackage{ifpdf}
\ifpdf % We are running pdfTeX in pdf mode
\documentclass[pdftex]{sigma}
\else
\documentclass{sigma}
\fi

\def\downindex#1{{}_{#1}}

\def\mathtext#1{\hbox{\rm{#1}}}

\def\parderop#1{\partial/\partial{#1}}

\def\D#1{D\downindex{#1}}

\def\X#1{{\bf X}_{#1}}
\def\prX#1{{\bf X}_{#1}^{(1)}}
\def\Y#1{{\bf Y}_{#1}}

\begin{document}

\allowdisplaybreaks

\renewcommand{\thefootnote}{$\star$}

\renewcommand{\PaperNumber}{066}

\FirstPageHeading

\ShortArticleName{Exact Solutions by the Method of Group Foliation Reduction}

\ArticleName{Exact Solutions of Nonlinear Partial \\  Dif\/ferential Equations
by the Method\\ of Group Foliation Reduction\footnote{This paper is a
contribution to the Special Issue ``Symmetry, Separation, Super-integrability and Special Functions~(S$^4$)''. The
full collection is available at
\href{http://www.emis.de/journals/SIGMA/S4.html}{http://www.emis.de/journals/SIGMA/S4.html}}}

\Author{Stephen C. ANCO~$^\dag$, Sajid ALI~$^\ddag$ and Thomas WOLF~$^\dag$}

\AuthorNameForHeading{S.C.~Anco, S.~Ali and T.~Wolf}

\Address{$^\dag$~Department of Mathematics, Brock University, St. Catharines, ON L2S 3A1 Canada}
\EmailD{\href{mailto:sanco@brocku.ca}{sanco@brocku.ca}, \href{mailto:twolf@brocku.ca}{twolf@brocku.ca}}

\Address{$^\ddag$~School of Electrical Engineering and Computer Sciences,\\
\hphantom{$^\ddag$}~National University of Sciences and Technology,
H-12 Campus, Islamabad 44000, Pakistan}
\EmailD{\href{mailto:sajid_ali@mail.com}{sajid{\_}ali@mail.com}}

\ArticleDates{Received March 05, 2011, in f\/inal form July 03, 2011;  Published online July 12, 2011}

\Abstract{A novel symmetry method
for f\/inding exact solutions to nonlinear PDEs
is illustrated by applying it to a class of semilinear reaction-dif\/fusion equations.
The method uses a separation ansatz to solve
an equivalent f\/irst-order group foliation system
whose independent and dependent variables
respectively consist of the invariants and dif\/ferential invariants of
a given one-dimensional group of point symmetries
for the class of reaction-dif\/fusion equations.
With this group-foliation reduction method,
solutions of the reaction-dif\/fusion equations
are obtained in an explicit form, including
group-invariant similarity solutions and travelling-wave solutions,
as well as dynamically interesting solutions that are not invariant under
any of the point symmetries admitted by the equations in this class.}


\Classification{35K58; 35C06; 35A25; 58J70; 34C14}

\section{Introduction}

The construction of group foliations using admitted point symmetry groups
for nonlinear partial dif\/ferential equations (PDEs)
is originally due to Lie and Vessiot
and was revived in its modern form by Ovsiannikov \cite{Ovsiannikov}.
In general a group foliation converts a given nonlinear PDE
into an equivalent f\/irst-order PDE system,
called the group-resolving equations,
whose independent and dependent variables respectively consist of
the invariants and dif\/ferential invariants of
a given one-dimensional group of point symmetry transformations.
Each solution of the group-resolving equations geometrically corresponds to
an explicit one-parameter family of exact solutions of
the original nonlinear PDE, such that the family is closed under
the given one-dimensional symmetry group
acting in the solution space of the PDE.

Because a group foliation contains all solutions of the given nonlinear PDE,
ansatzes or dif\/ferential-algebraic constraints must be used to reduce
the group-resolving equations into an overdetermined system
for the purpose of obtaining explicit solutions.
Compared with classical symmetry reduction \cite{Olver-book,BlumanAnco-book},
a main dif\/f\/iculty to-date has been how to f\/ind
ef\/fective, systematic ansatzes that lead to useful reductions.

An important step toward overcoming this dif\/f\/iculty has been taken
in recent work \cite{AncoLiu,AncoAliWolf}
on f\/inding exact solutions to semilinear wave equations and heat equations
with power nonlinearities.
Specif\/ically, this work demonstrates that the group-resolving equations
for such nonlinear PDEs have solutions arising from a simple separation ansatz
in terms of the group-invariant variables.
Through this ansatz,
many explicit solutions to the nonlinear PDE are easily found,
whose form would not be readily obvious just by trying simple direct ansatzes
using the original independent and dependent variables in the nonlinear PDE,
or by simply writing down the form for classical group-invariant solutions.
In particular,
some of these solutions are not invariant under
any of the point symmetries of the nonlinear PDE
and thus fall completely outside of classical symmetry reduction
(while others coincide with explicit group-invariant solutions).
Most importantly for applications,
many of the solutions also have interesting analytical properties
related to critical dynamics, blow-up behaviour, asymptotic behaviour and attractors.

We will illustrate this group-foliation reduction method
by applying it to obtain explicit exact solutions for
the class of semilinear reaction-dif\/fusion equations
\begin{gather}\label{reactdiffeq}
u_{t} = u_{rr} + m r^{-1}u_{r} + (p-k u^q)u,
\qquad
k={\rm const}\neq0,
\qquad
p={\rm const}
\end{gather}
for $u(t,r)$, with a nonlinearity power $q\neq 0,-1$,
and a spatial derivative coef\/f\/icient $m$.
When $m$ is a positive integer,
the equation \eqref{reactdiffeq} physically models radial dif\/fusion
with a nonlinear reaction source/sink term in $m+1$ spatial dimensions,
where $r$ is the radial coordinate.
When $m$ is zero, this equation similarly models reaction-dif\/fusion
in one spatial dimension,
where $r$ denotes the full-line coordinate.
Finally, when $m$ is not a non-negative integer,
the equation \eqref{reactdiffeq} instead is a model for two-dimensional
radial reaction-dif\/fusion with an extra spatial-derivative reaction term
$(1-m)r^{-1} u_r$ which can be shown \cite{AncoAliWolf}
to add a source/sink contribution $(1-m)\lim\limits_{r\rightarrow 0}u$
to the net f\/low integral $\frac{d}{dt}\int_0^\infty u rdr$
in two spatial dimensions.

Note that the constants $p$, $k$ in equation \eqref{reactdiffeq}
can be freely scaled by a two-parameter transformation
$t\rightarrow {\lambda_1}^2 t$, $r\rightarrow \lambda_1 r$, $u\rightarrow \lambda_2 u$,
under which $p\rightarrow {\lambda_1}^2 p$, $k\rightarrow {\lambda_2}^q k$
(with $\lambda_1,\lambda_2\neq0$).
Thus there is no loss of generality in assuming $k=\pm 1$ and $p=0,\pm 1$.

The symmetry structure of this class of reaction-dif\/fusion equations \eqref{reactdiffeq}
is given by \cite{Dorodnitsyn}
\begin{alignat}{3}
&\mathtext{ time translation } \quad &&
\X{1} =\parderop{t} \quad\mathtext{ for all $m,\,q,\,p$}, &
\label{ttranssymm}\\
&\mathtext{ scaling } \quad &&
\X{2} =2t\parderop{t} + r\parderop{r} -(2/q) u\parderop{u}
\quad\mathtext{ only for $p=0$}, &
\label{scalsymm}\\
&\mathtext{ space translation } \quad &&
\X{3} =\parderop{r} \quad\mathtext{ only for $m=0$},&
\label{rtranssymm}
\end{alignat}
where $\X{}$ is the inf\/initesimal generator of a one-parameter group of
point transformations acting on $(t,r,u)$.
For constructing a group foliation,
it is natural to use the time translation generator~\eqref{ttranssymm},
since this is the only point symmetry admitted for
all cases of the parameters $q$, $m$ and all values of the constants $p$, $k$.

In Section~\ref{method},
we f\/irst set up the system of group-resolving equations
given by the time-translation symmetry~\eqref{ttranssymm}
for the reaction-dif\/fusion equation~\eqref{reactdiffeq},
which uses the invariants and dif\/ferential invariants of
the symmetry generator $\X{1}$
as the independent and dependent variables in the system.
We next state the form required for solutions of the group-resolving system
to correspond to group-invariant solutions of
the reaction-dif\/fusion equation~\eqref{reactdiffeq}
with respect to the point symmetries generated by $\X{1}$, $\X{2}$, $\X{3}$.

In Section~\ref{results},
we explain the separation ansatz for directly reducing
the system of group-resolving equations.
This reduction yields an overdetermined system of dif\/ferential-algebraic equations
which can be readily solved by computer algebra.
We present the explicit solutions of these equations
and then we derive the resulting exact solutions of the reaction-dif\/fusion equation.
These solutions include explicit similarity solutions in the case $p=0$,
and explicit travelling wave solutions
in addition to explicit non-invariant solutions in the case $m=0$.

In Section~\ref{subspaces},
we show how the success of the reduction ansatz
can be understood equivalently as constructing
partially-invariant subspaces for a nonlinear operator that arises
in a natural way from the structure of the group-resolving equations.
This important observation puts our method on a wider mathematical foundation
within the general theory of invariant subspaces developed by
Galaktionov~\cite{Galaktionov-book}.

Finally, we make some general concluding remarks in Section~\ref{remarks}.

\section{Group-resolving equations and symmetries}
\label{method}

To proceed with setting up the time-translation group foliation
for the reaction-dif\/fusion equation~\eqref{reactdiffeq},
we f\/irst write down the invariants (in terms of $t$, $r$, $u$)
\begin{gather}
x=r ,\qquad
v=u ,
\label{xv}
\end{gather}
satisfying $\X{1}x=\X{1}v=0$,
and the dif\/ferential invariants (in terms of $u_t$, $u_r$)
\begin{gather}
G= u_{t} ,\qquad
H= u_{r} ,
\label{GH}
\end{gather}
satisfying $\prX{1}G=\prX{1}H=0$ where
$\prX{1}$ is the f\/irst-order prolongation of the generator \eqref{ttranssymm}.
Here $x$ and $v$ are mutually independent,
while $G$ and $H$ are related by equality of mixed~$r$,~$t$ derivatives
on $u_{t}$ and $u_{r}$, which gives
\begin{gather}
\D{r} G = \D{t} H,
\label{GHmixedeq}
\end{gather}
where $\D{r}$, $\D{t}$ denote total derivatives with respect to $r$, $t$.
In addition,
$v$, $G$, $H$ are related through the reaction-dif\/fusion equation~\eqref{reactdiffeq}
by
\begin{gather}
G -r^{-m}\D{r}\big(r^{m}H\big) = \big(p-kv^q\big)v .
\label{GHreactdiffeq}
\end{gather}
Now we put $G=G(x,v)$, $H=H(x,v)$ into equations \eqref{GHmixedeq} and \eqref{GHreactdiffeq}
and use equation~\eqref{xv} combined with the chain rule
to arrive at a f\/irst-order PDE system
\begin{gather}
G_x + HG_v -GH_v =0 ,
\label{GHeq1}\\
G - m H/x -H_x -HH_v =\big(p-kv^q\big)v ,
\label{GHeq2}
\end{gather}
with independent variables $x$, $v$, and dependent variables $G$, $H$.
These PDEs are called the {\it time-translation-group resolving system}
for the reaction-dif\/fusion equation~\eqref{reactdiffeq}.

The respective solution spaces of
equation~\eqref{reactdiffeq} and system \eqref{GHeq1}, \eqref{GHeq2}
are related by a group-invariant mapping that is def\/ined through
the invariants~\eqref{xv} and dif\/ferential invariants~\eqref{GH}.

\begin{lemma}
Solutions $(G(x,v),H(x,v))$ of
the time-translation-group resolving system \eqref{GHeq1}, \eqref{GHeq2}
are in one-to-one correspondence with
one-parameter families of solutions $u(t,r,c)$ of
the reac\-tion-diffu\-sion equation \eqref{reactdiffeq}
satisfying the translation-invariance property
\begin{gather*}
u(t+\epsilon,r,c) = u(t,r,\tilde c(\epsilon,c)),
%\label{orbit}
\end{gather*}
where $\tilde c(0,c)=c$ in terms of
an arbitrary constant $c$ and parameter $\epsilon$,
such that
\begin{gather}
u_{t} =G(r,u) ,\qquad
u_{r} =H(r,u)
\label{umap}
\end{gather}
constitutes a consistent pair of parametric first-order ODEs
whose integration constant is~$c$.
\end{lemma}

We now examine the relationship between the symmetry structure of
the reaction-dif\/fusion equation~\eqref{reactdiffeq}
and the symmetry structure inherited by
the time-translation-group resolving system \eqref{GHeq1}, \eqref{GHeq2}.

Firstly,
through the identif\/ications def\/ined by the variables \eqref{xv}, \eqref{GH},
the prolongation of any point symmetry generator
$\X{}=a_1\X{1}+ a_2\X{2}+ a_3\X{3}$ of equation~\eqref{reactdiffeq}
has a natural projection to a point symmetry generator
$\Y{}=a_2\prX{2}+ a_3\prX{3}$ modulo $\prX{1}$ of system \eqref{GHeq1}, \eqref{GHeq2}.
The time-translation $\X{1}$ thus gets annihilated by this projection,
i.e.~$\Y{1}=0$,
while the scaling $\X{2}$ and the space-translation $\X{3}$
respectively project to
\begin{gather}
\Y{2} = x\parderop{x} -(2/q) v\parderop{v} -2(1+1/q)G\parderop{G} - (1+2/q)H\parderop{H}
\qquad\mathtext{when $p=0$},
\label{GHscalsymm}
\end{gather}
and
\begin{gather}
\Y{3} =\parderop{x}
\qquad\mathtext{when $m=0$}.
\label{GHxtranssymm}
\end{gather}

Secondly,
with respect to these inherited symmetries \eqref{GHscalsymm} and \eqref{GHxtranssymm},
the system \eqref{GHeq1}, \eqref{GHeq2} has a reduction to ODEs yielding solutions
such that $(G,H)$ is invariant respectively under scalings
\begin{gather*}
x\rightarrow \lambda x, \qquad
v\rightarrow \lambda^{-2/q} v
\qquad\mathtext{when $p=0$},
\end{gather*}
and under translations
\begin{gather*}
x\rightarrow x+\epsilon
\qquad\mathtext{when $m=0$}.
\end{gather*}
Thus,
translation-invariant solutions have the form
\begin{gather}
(G,H) = (g(v),h(v))
\label{transinvGH}
\end{gather}
satisfying the ODE system
\begin{gather}
(h/g)'=0 ,\qquad
g-hh' =(p-kv^q)v.
\label{transinvGHeqs}
\end{gather}
Scaling-invariant solutions have the form
\begin{gather}
(G,H) = (x^{-2-2/q}g(V),x^{-1-2/q}h(V)),
\qquad
V=v x^{2/q}
\label{scalinvGH}
\end{gather}
satisfying the ODE system
\begin{gather}
((h+2V/q)/g)'=-2/g ,\qquad
g+(1-m+2/q)h-(h+2V/q)h' =-kV^{q+1}.
\label{scalinvGHeqs}
\end{gather}
Integration of the parametric ODEs~\eqref{umap} for such solutions
\eqref{transinvGH}, \eqref{transinvGHeqs} and \eqref{scalinvGH}, \eqref{scalinvGHeqs}
leads to the following two correspondence results.

\begin{lemma}\label{travellingwaveform}
In the case $m=0$,
there is a one-to-one correspondence between solutions of
the translation-group resolving system~\eqref{GHeq1},~\eqref{GHeq2}
with the invariant form \eqref{transinvGH}
and one-parameter families of travelling-wave solutions of
the reaction-diffusion equation~\eqref{reactdiffeq}
given by the group-invariant form
$u=f(\xi)$ where, modulo time-translations $t\rightarrow t+c$,
the variable $\xi=r-t/a$ is an invariant of the translation symmetry
$\X{}= a\parderop{t} + \parderop{r} =a\X{1}+\X{3}$
in terms of some constant $a$
$($determined by ODE system~\eqref{transinvGHeqs}$)$.
\end{lemma}

\begin{lemma}\label{similarityform}
In the case $p=0$,
there is a one-to-one correspondence between solutions of
the translation-group resolving system \eqref{GHeq1}, \eqref{GHeq2}
with the invariant form~\eqref{scalinvGH}
and one-parameter families of similarity solutions of
the reaction-diffusion equation~\eqref{reactdiffeq}
given by the group-invariant form
$u=r^{-2/q}f(\xi)$ where, modulo time-translations $t\rightarrow t+c$,
the variable $\xi=t/r^2$ is an invariant of the scaling symmetry
$\X{} = 2t\parderop{t} + r\parderop{r} -(2/q) u\parderop{u}=\X{2}$.
\end{lemma}

Furthermore, in all cases,
static solutions $u(r)$ of the reaction-dif\/fusion equation
correspond to solutions of the translation-group resolving system with $G=0$.
Hereafter we will be interested only in solutions with $G\neq0$ and $H\neq0$,
corresponding to dynamical solutions $u(t,r)$
of the reaction-dif\/fusion equation.

\section{Main results}
\label{results}

To f\/ind explicit solutions of
the group foliation system \eqref{GHeq1}, \eqref{GHeq2}
for $(G(x,v),H(x,v))$,
we will make use of the same general homogeneity features
utilized in~\cite{AncoLiu,AncoAliWolf}.
First,
the non-derivative terms $(p-k u^q)u$
in the reaction-dif\/fusion equation~\eqref{reactdiffeq}
appear only as an inhomogeneous term in equation~\eqref{GHeq2}.
Second,
in both equations \eqref{GHeq1} and \eqref{GHeq2}
the linear terms involve no derivatives with respect to~$v$.
Third,
the nonlinear terms in the homogeneous equation~\eqref{GHeq1}
have the skew-symmetric form $HG_v-GH_v$,
while the only nonlinear term appearing in the non-homogeneous equation~\eqref{GHeq2}
has the symmetric form $HH_v=(H^2/2)_v$.
Based on these features, this system can be expected to have solutions
given by the separable power form
\begin{gather}
G=g_1(x) v + g_2(x) v^a,\qquad
H=h_1(x) v + h_2(x) v^a,\qquad
a\neq 1 .
\label{GHansatz}
\end{gather}

For such a separation ansatz~\eqref{GHansatz},
the linear terms $G_x$, $G$, $H/x$, $H_x$
in equations \eqref{GHeq1} and \eqref{GHeq2} will contain the same powers $v$, $v^a$
that appear in both $G$ and $H$,
and moreover the nonlinear term $HG_v-H_vG$
in the homogeneous equation~\eqref{GHeq1} will produce only the power~$v^a$
due to the identities $v^a(v)_v- (v^a)_v v= (a-1)v^a$
and $v(v)_v- (v)_v v= v^a(v^a)_v- (v^a)_v v^a = 0$.
Thus, equation~\eqref{GHeq1} can be satisf\/ied by having the coef\/f\/icients of
$v$ and $v^a$ separately vanish.
Similarly the nonlinear term $HH_v$
in the non-homogeneous equation~\eqref{GHeq2} will only yield the powers
$v$, $v^a$, $v^{2a-1}$.
Since we have $a\neq1$ and $q\neq0$,
equation~\eqref{GHeq2} can be satisf\/ied by again having the coef\/f\/icients of~$v$ and~$v^a$ separately vanish
and by also having the term containing $v^{2a-1}$
balance the inhomogeneous term $k v^{q+1}$.
In this fashion we f\/ind that equations~\eqref{GHeq1} and~\eqref{GHeq2}
reduce to an overdetermined system of 5 algebraic-dif\/ferential equations
for $g_1(x)$, $g_2(x)$, $h_1(x)$, $h_2(x)$,
together with the relation $a=1+q/2$.
This system can be solved by a systematic integrability analysis.
We have carried out this analysis using the computer algebra program
{\sc Crack}~\cite{crack},
which contains a wide repertoire of modules for
reduction of dif\/ferential order and polynomial degree,
splittings with respect to the independent variable(s),
eliminations, substitutions, factorizations, integrations,
and length-shortening of equations, among others.

\begin{proposition}
For $q\neq 0,-1$,
the separation ansatz \eqref{GHansatz} yields altogether
six solutions of the translation-group resolving system \eqref{GHeq1}, \eqref{GHeq2}
with $G\neq 0$ and $H\neq0$:
\begin{gather}
G= \pm (3-m) \left(\frac{m-1}{m-2}k\right)^{1/2} v^{(m-2)/(m-1)}/x,
\nonumber\\
H= \pm \left(\frac{m-1}{m-2}k\right)^{1/2} v^{(m-2)/(m-1)} +(1-m)v/x,
\nonumber\\\qquad
m\neq 1,2, \qquad q=2/(1-m), \qquad p=0, \qquad k\gtrless 0 \quad((m-1)(m-2)\gtrless 0);
\label{sol3}\\
G = \frac{q+4}{q+2} \big(pv+ (pk)^{1/2} v^{1+q/2}\big),
\nonumber\\
H = \pm \left( \left(\frac{2p}{q+2}\right)^{1/2} v
+ \left(\frac{2k}{q+2}\right)^{1/2} v^{1+q/2} \right),
\nonumber\\\qquad
m=0,\qquad q\gtrless -2, \qquad p\gtrless 0, \qquad k\gtrless 0;
\label{sol1a}\\
G = \frac{q+4}{q+2} \big(pv -(pk)^{1/2} v^{1+q/2}\big),
\nonumber\\
H = \pm \left( \left(\frac{2p}{q+2}\right)^{1/2} v
- \left(\frac{2k}{q+2}\right)^{1/2} v^{1+q/2} \right),
\nonumber\\\qquad
m=0,\qquad q\gtrless -2, \qquad p\gtrless 0, \qquad k\gtrless 0;
\label{sol1b}\\
G= \frac{3}{2}\big(pv \pm (pk) ^{1/2} \tanh \big((p/2)^{1/2}(x+c_1)\big) v^2\big),
\nonumber\\
H= \pm (k/2)^{1/2} v^2+ (p/2)^{1/2} \tanh \big((p/2)^{1/2}(x+c_1)\big) v,
\nonumber\\\qquad
m=0, \qquad q=2, \qquad p>0, \qquad k>0;
\label{sol2a}\\
G= \frac{3}{2}\big( pv \mp (-pk )^{1/2}\tan \big((-p/2)^{1/2}(x+c_1)\big) v^2 \big),
\nonumber\\
H= \pm (k/2)^{1/2} v^2 - (-p/2)^{1/2} \tan \big((-p/2)^{1/2}(x+c_1)\big) v,
\nonumber\\\qquad
m=0, \qquad q=2, \qquad p<0, \qquad k>0;
\label{sol2b}\\
G= \pm 3(k/2)^{1/2} v^2/(x+c_1),
\nonumber\\
H= \pm (k/2)^{1/2} v^2 +v/(x+c_1),
\nonumber\\\qquad
m=0, \qquad q=2, \qquad p=0, \qquad k>0.
\label{sol4}
\end{gather}
Solution \eqref{sol3} satisfies
the scaling-invariance reduction~\eqref{scalinvGH};
solutions~\eqref{sol1a} and~\eqref{sol1b}
satisfy the translation-invariance reduction~\eqref{transinvGH};
solution~\eqref{sol4} for $c_1=0$ satisfies
the scaling-invariance reduction~\eqref{scalinvGH}.
\end{proposition}

\begin{remark}
A shift of the arbitrary constant
$c_1\rightarrow c_1 \pm i \pi/(2p)^{1/2}$ in solution \eqref{sol2a}
and $c_1\rightarrow c_1 \pm \pi/(-2p)^{1/2}$ in solution \eqref{sol2b}
respectively yields the two further solutions
\begin{gather*}
G= \frac{3}{2}\big(pv \pm (pk) ^{1/2} \coth\big((p/2)^{1/2}(x+c_1)\big) v^2\big ),
\nonumber\\
H= \pm (k/2)^{1/2} v^2+ (p/2)^{1/2} \coth\big((p/2)^{1/2}(x+c_1)\big) v,
\nonumber\\\qquad
m=0, \qquad q=2, \qquad p>0, \qquad k>0,
%\label{sol2c}
\end{gather*}
and
\begin{gather*}
G= \frac{3}{2}\big( pv \pm (-pk )^{1/2}\cot\big((-p/2)^{1/2}(x+c_1)\big) v^2 \big),
\nonumber\\
H= \pm (k/2)^{1/2} v^2 + (-p/2)^{1/2} \cot\big((-p/2)^{1/2}(x+c_1)\big) v,
\nonumber\\\qquad
m=0, \qquad q=2, \qquad p<0, \qquad k>0,
%\label{sol2d}
\end{gather*}
as obtained through the trigonometric identities
\begin{gather*}
\tanh(\theta \pm i \pi/2) = \coth(\theta),
\qquad
\tan(\theta \pm \pi/2) = -\cot(\theta).
\end{gather*}
\end{remark}

The form of the separation ansatz \eqref{GHansatz} can be naturally generalized
to include additional powers
\begin{gather}
G=g_1(x) v + \sum_{i=1}^{N} g_{1+i}(x) v^{a_i} ,\qquad
H=h_1(x) v + \sum_{i=1}^{N} h_{1+i}(x) v^{a_i} ,
\label{generalGHansatz}
\end{gather}
where
\begin{gather}
a_i\neq 1,\qquad
a_i-a_j\neq 0 .
\label{GHconds}
\end{gather}
This multi-term ansatz (for $N\geq 2$) leads to
a considerably more complicated analysis
compared to the previous two-term ansatz (with $N=1$).
Specif\/ically,
the homogeneous equation~\eqref{GHeq1} now contains the powers
$v^{a_i+a_j-1}$ in addition to $v$, $v^{a_i}$,
while the non-homogeneous equation~\eqref{GHeq2} contains the further powers~$v^{2a_i-1}$, and~$v^{q+1}$.
To determine the exponents in these powers,
a~systematic examination of the possible balances
(which rapidly increase in number with $N$) is necessary.
We carry out this analysis by computer algebra,
again using the program {\sc Crack}~\cite{crack}.
In particular, for any f\/ixed $N\geq 1$,
{\sc Crack} can be run automatically
by setting up a priority list of modules to be tried recursively
with case splittings given a high priority,
in order to split equations \eqref{GHeq1} and \eqref{GHeq2} under the ansatz
\eqref{generalGHansatz}, \eqref{GHconds}
and in each case solve the resulting overdetermined system of
algebraic-dif\/ferential equations for $g_i$ and $h_i$, $i=1,2,\ldots,N$.
Importantly,
the modules in {\sc Crack} are able to organize the case distinctions
in an ef\/f\/icient way that lessens a combinatorial explosion
by using algebraic conditions (equalities and inequalities)
arising in the steps of solving the equations
to keep track of whether a possible balance of two powers of~$v$
generates a new case (or subcase) which has to be solved or not.
Running {\sc Crack},
we f\/ind that the system
\eqref{GHeq1}, \eqref{GHeq2}, \eqref{generalGHansatz}, \eqref{GHconds}
has solutions when $N=2$ but not when $N=3$.
(Based on this outcome we have not tried to investigate the system
when $N>3$.)

\begin{proposition}
For $q\neq 0,-1$, and $N\leq 3$,
the separation ansatz~\eqref{generalGHansatz} yields four additional
solutions of the translation-group resolving system \eqref{GHeq1}, \eqref{GHeq2}
with $G\neq0$:
\begin{gather}
G= 3p\big( v+(k/p)^{2/3}v^{1/2} +(k/p)^{1/3} \big),
\nonumber\\
H= \pm(2p)^{1/2} \big( v+(k/p)^{2/3}v^{1/2} +(k/p)^{1/3} \big),
\nonumber\\\qquad
m=0, \qquad q=-3/2, \qquad p>0, \qquad k\gtrless0;
\label{scalingsol5}
\\
G= 4p\big( v+(-4k/p)^{1/4}v^{1/3} +(-k/p)^{1/2}v^{-1/3} \big),
\nonumber\\
H= \pm(3p)^{1/2} \big( v+(-4k/p)^{1/4}v^{1/3} +(-k/p)^{1/2}v^{-1/3} \big),
\nonumber\\\qquad
m=0,\qquad q=-8/3, \qquad p>0, \qquad k<0;
\label{scalingsol6a}
\\
G= 4p\big( v -(-4k/p)^{1/4}v^{1/3} +(-k/p)^{1/2}v^{-1/3} \big),
\nonumber\\
H= \pm(3p)^{1/2} \big( v -(-4k/p)^{1/4}v^{1/3} +(-k/p)^{1/2}v^{-1/3} \big),
\nonumber\\\qquad
m=0,\qquad q=-8/3, \qquad p>0, \qquad k<0;
\label{scalingsol6b}
\\
G = \pm 3(-k/8)^{1/2}\big( v \pm (-2k)^{-1/2} x^{-1} \big)^2/x ,
\nonumber\\
H = \pm (-k/2)^{1/2}\big( v^2 + 3/(2k) x^{-2} \big),
\nonumber\\\qquad
m=3/2, \qquad q=2, \qquad p=0, \qquad k<0.
\label{scalingsol7}
\end{gather}
Solutions \eqref{scalingsol5}, \eqref{scalingsol6a}, \eqref{scalingsol6b}
satisfy the translation-invariance reduction \eqref{transinvGH};
solution \eqref{scalingsol7} satisfies
the scaling-invariance reduction~\eqref{scalinvGH}.
\end{proposition}

We now obtain explicit solutions $u(t,r)$ of
the reaction-dif\/fusion equation \eqref{reactdiffeq}
from the solutions $(G(x,v),H(x,v))$
of its translation-group resolving system \eqref{GHeq1}, \eqref{GHeq2}
by integrating the corresponding pair of
parametric f\/irst-order ODEs~\eqref{umap}.
This integration yields a one-parameter solution family~$u(t,r,c)$
which is closed under the action of the group of time-translations
$t\rightarrow t+\epsilon$.

\begin{theorem}
The semilinear reaction-diffusion equation \eqref{reactdiffeq}
has the following exact solutions
arising from the explicit solutions of
its translation-group resolving system
found in Propositions~{\rm 1} and~{\rm 2} $($and Remark~$1)$:
\begin{gather}
u =\left ( \pm \left ( \frac{k}{(m-2)(m-1)}\right )^{1/2} \left(\frac{r}{2}-\frac{(m-3)(t+c)}{r}\right)\right) ^{m-1},
\nonumber\\\quad
m\neq 1,2, \qquad
q=2/(1-m), \qquad
p=0, \qquad
k\gtrless 0 \quad((m-1)(m-2)\gtrless 0);
\label{usol1}
\\
u= \left (- \left ( k/p \right ) ^{1/2}+\exp{\left (\mp q \left ( \frac{p}{2(q+2)} \right ) ^{1/2}
 \left ( r\pm (q+4) \left ( \frac{p}{2(q+2)} \right )^{1/2}(t+c) \right )\right )}   \right )^{-2/q} ,
\nonumber\\\quad
m=0, \qquad
q\gtrless -2, \qquad
p\gtrless 0, \qquad
k\gtrless 0;
\label{usol2}
\\
u= \left ( \left ( k/p \right )^{1/2} +\exp{\left (\mp q  \left (\frac{p}{2(q+2)} \right ) ^{1/2}
\left ( r\pm (q+4) \left ( \frac{p}{2(q+2)} \right ) ^{1/2}(t+c) \right )\right )} \right )^{-2/q} ,
\nonumber\\\quad
m=0, \qquad
q\gtrless -2, \qquad
p\gtrless 0, \qquad
k\gtrless 0;
\label{usol3}
\\
u = \frac{\cosh( (p/2)^{1/2}(r+c_1) )}
{(k/p)^{1/2} \sinh( (p/2)^{1/2}(r+c_1) ) \pm \exp(-3p(t+c)/2)},
\nonumber\\\quad
m=0, \qquad
q=2, \qquad
p>0, \qquad
k>0;
\label{usol6a}
\\
u = \frac{\sinh( (p/2)^{1/2}(r+c_1) )}
{(k/p)^{1/2} \cosh( (p/2)^{1/2}(r+c_1) ) \pm \exp(-3p(t+c)/2)},
\nonumber\\\quad
m=0, \qquad
q=2, \qquad
p>0, \qquad
k>0;
\label{usol6b}
\\
u= \frac{\cos( (-p/2)^{1/2}(r+c_1) )}
{(-k/p)^{1/2} \sin( (-p/2)^{1/2}(r+c_1) ) \pm \exp(-3p(t+c)/2)},
\nonumber\\\quad
m=0, \qquad
q=2, \qquad
p<0, \qquad
k>0;
\label{usol7a}
\\
u= \frac{\sin( (-p/2)^{1/2}(r+c_1) )}
{(-k/p)^{1/2} \cos( (-p/2)^{1/2}(r+c_1) ) \mp \exp(-3p(t+c)/2)},
\nonumber\\\quad
m=0, \qquad
q=2, \qquad
p<0, \qquad
k>0;
\label{usol7b}
\\
u =\left ( \pm (k/2)^{1/2} \left(\frac{r+c_1}{2}-\frac{3(t+c)}{r+c_1}\right)\right)^{-1},
\nonumber\\\quad
m=0,\qquad
q=2, \qquad
p=0, \qquad
k>0;
\label{usol8}
\\
u=\frac{t+c-r^2}{\sqrt{2k}(t+c+r^2/3)r},
\nonumber\\\quad
m= 3/2,\qquad
q=2, \qquad
p=0, \qquad
k>0;
\label{usol9}
\\
u= (k/(8p))^{2/3}\left( \sqrt{3}\cot\phi(\xi) -1 \right)^2
\quad\mathtext{where}\quad
\xi = r \pm 3(p/2)^{1/2}(t+c),
\nonumber\\
\exp(-\phi(\xi)/\sqrt{3})\sin\phi(\xi) = \sqrt{3}(k/(8p))^{1/3}\exp\big((2p)^{1/2}\xi\big),
\nonumber\\\quad
m=0, \qquad
q=-3/2, \qquad
p>0, \qquad
k\gtrless 0;
\label{usol4}
\\
u= (-k/(4p))^{3/8}\left( \cot\phi(\xi) -1 \right)^{3/2}
\quad\mathtext{where}\quad
\xi = r \pm 4(p/3)^{1/2}(t+c),
\nonumber\\
\exp(-2\phi(\xi))\sin\phi(\xi) = (-k/(4p))^{1/4}\exp\big(2(p/3)^{1/2}\xi\big),
\nonumber\\ \quad
m=0, \qquad
q=-8/3, \qquad
p>0, \qquad
k<0;
\label{usol5a}
\\
u= (-k/(4p))^{3/8}\left( \cot\phi(\xi) +1 \right)^{3/2}
\quad\mathtext{where}\quad
\xi = r \pm 4(p/3)^{1/2}(t+c),
\nonumber\\
\exp(2\phi(\xi))\sin\phi(\xi) = (-k/(4p))^{1/4}\exp\big(2(p/3)^{1/2}\xi\big),
\nonumber\\ \quad
m=0, \qquad
q=-8/3, \qquad
p>0, \qquad
k<0,
\label{usol5b}
\end{gather}
where $c$, $c_1$ are arbitrary constants.
\end{theorem}

\begin{remark}
Solutions \eqref{usol6a} and \eqref{usol6b} are related by
shifts of the arbitrary constants
$c_1\rightarrow c_1 \pm i \pi/(2p)^{1/2}$
and $c\rightarrow c \mp i \pi/(3p)$
through the trigonometric identities
\begin{gather*}
\sinh(\theta\pm i \pi/2) = \pm i \cosh(\theta), \qquad
\cosh(\theta\pm i \pi/2) = \pm i \sinh(\theta),
\end{gather*}
while solutions \eqref{usol7a} and \eqref{usol7b} are related
in a similar way through the trigonometric identities
\begin{gather*}
\sin(\theta\pm  \pi/2) = \pm  \cos(\theta), \qquad
\cos(\theta\pm  \pi/2) = \mp  \sin(\theta).
\end{gather*}
\end{remark}

Modulo time-translations,
solutions~\eqref{usol1} and~\eqref{usol9} are similarity solutions,
since they have the form shown in Lemma~\ref{similarityform}.
These solutions have been obtained in our previous work~\cite{AncoAliWolf}
using a~two-term separation ansatz to solve the group-resolving equations
given by the scaling symmetry~\eqref{scalsymm}
for the reaction-dif\/fusion equation~\eqref{reactdiffeq} in the case $p=0$.
Solution~\eqref{usol8} can be obtained from the $m=0$ case of
solution~\eqref{usol1} after a space-translation is applied.

Solutions \eqref{usol2}, \eqref{usol3}, \eqref{usol4}, \eqref{usol5a} and \eqref{usol5b}
are travelling-wave solutions
with the form shown in Lemma~\ref{travellingwaveform}
(where the constant $a$ is the wave speed).
The two solutions \eqref{usol2} and \eqref{usol3}
were f\/irst obtained in~\cite{Vijayakumar}
through standard symmetry reduction.
To the best knowledge of the authors,
the other three solutions~\eqref{usol4}, \eqref{usol5a} and~\eqref{usol5b}
are new.

In contrast,
solutions \eqref{usol6a} to \eqref{usol7b} are non-invariant solutions,
as they do not have the form of travelling waves
whereas the symmetry group of the reaction-dif\/fusion equation~\eqref{reactdiffeq}
in the case $p\neq 0$ and $m=0$ is generated entirely by
time-translation and space-translation sym\-metries~\eqref{ttranssymm} and~\eqref{rtranssymm}.
These solutions have been found previously
in \cite{Clarkson, Arrigo} using Bluman and Cole's nonclassical method.

\section{Invariance properties of the reduction ansatz}
\label{subspaces}

As we will now explain,
the success of the separation ansatzes \eqref{GHansatz} and \eqref{generalGHansatz}
for reducing the system of group-foliation equations \eqref{GHeq1} and \eqref{GHeq2}
has a mathematically natural interpretation
within the theory of invariant subspaces \cite{Galaktionov-book}.

We start by writing the group-foliation equations in an
evolutionary operator form
\begin{gather}
\begin{pmatrix}
G \\ H
\end{pmatrix}_{x}
= \Phi \left(
\begin{pmatrix}
G \\ H
\end{pmatrix}
\right)
=
\begin{pmatrix}
0 \\ (k v^{q}-p)v
\end{pmatrix}
+
\begin{pmatrix}
0 \\ G  -mH/x
\end{pmatrix}
+
\begin{pmatrix}
G H_v-H G_v \\ -H H_v
\end{pmatrix},
\label{GHevoleq}
\end{gather}
where $\Phi$ def\/ines a nonlinear operator acting on
the pair of variables $(G,H)$.
Then we view the ansatz \eqref{GHansatz}
as def\/ining a linear space of functions with a separable power form
\begin{gather}
\begin{pmatrix}
G \\ H
\end{pmatrix}
=
\begin{pmatrix}
g_{1} \\ h_{1}
\end{pmatrix} v
+
\begin{pmatrix}
g_{2} \\ h_{2}
\end{pmatrix}
v^{a},
\qquad a\neq 1,
\label{GHlinspace}
\end{gather}
where the coef\/f\/icients depend only on $x$.
This linear space \eqref{GHlinspace} is not invariant under $\Phi$
because
\begin{gather*}
\Phi \left(
\begin{pmatrix} g_{1} \\ h_{1} \end{pmatrix} v
+
\begin{pmatrix} g_{2} \\ h_{2} \end{pmatrix} v^{a}
\right)\nonumber\\
\qquad{}
=
\begin{pmatrix}
0 \\ g_{1} -m h_{1}/x -h_{1}^2 -p
\end{pmatrix}
v
+
\begin{pmatrix}
(a-1)(h_{2}g_{1}-h_{1}g_{2}) \\ g_{2} -m h_{2}/x-(a+1)h_{1}h_{2}
\end{pmatrix}
v^a
\nonumber\\
 \qquad\quad
{}+
\begin{pmatrix}
0 \\ -a h_{2}^2
\end{pmatrix}
v^{2a-1}
+
\begin{pmatrix}
0 \\ k
\end{pmatrix}
v^{q+1}
\end{gather*}
produces terms that have additional powers $v^{2a-1}$, $v^{q+1}$.

However, the operator $\Phi$ will preserve a set of functions
contained in the linear space~\eqref{GHlinspace}
if the power $a$ and the coef\/f\/icients $g_{1}$, $g_{2}$, $h_{1}$, $h_{2}$
satisfy the conditions
\begin{gather*}
2a-1 = q+1, \qquad ah_{2}^2 =k,
\end{gather*}
so that the terms containing $v^{2a-1}$ and  $v^{q+1}$ cancel out.
In this sense the linear space will then be partially-invariant~\cite{Galaktionov-book} under~$\Phi$.
As a consequence of this invariance property,
the evolution equation~\eqref{GHevoleq} for $(G,H)$ reduces to
an overdetermined system of dif\/ferential-algebraic equations
which we can solve for $g_{1}$, $g_{2}$, $h_{1}$, as discussed in Section~\ref{method}.

The separable power ansatz~\eqref{GHansatz} thus can be viewed
as explicitly constructing a partially-invariant linear subspace
for the group-foliation operator $\Phi$.
A similar discussion applies to the multi-term ansatz~\eqref{generalGHansatz}.

It is worth pointing out that
apart from the mathematical interpretation given to the an\-sat\-zes~\eqref{GHansatz} and~\eqref{generalGHansatz},
the theory of (partially) invariant linear subspaces
does not provide any general constructive method or algorithm
for f\/inding a successful ansatz.
So the main aspect of the present work is to demonstrate
the ef\/fectiveness of a general separation ansatz for solving
group-foliation equations.
We also emphasize that this approach is much more ef\/fective than
if an analogous separation method were to be applied
either directly for solving the given reaction-dif\/fusion PDE
or instead for seeking a (partially) invariant linear subspace
for that PDE itself.

\section{Concluding remarks}
\label{remarks}

In general the method of group foliation reduction using a separation ansatz
as illustrated in this paper is able to yield
exact solutions to nonlinear 2nd order PDEs with power nonlinearities.

This method works with any admitted group of point (or contact) symmetries
and gives a~systematic reduction of the group foliation equations
into an overdetermined system consisting of algebraic equations and
1st order dif\/ferential equations
that can be derived and in most cases solved
by means of computer algebra
(e.g.\ using the program {\sc Crack} \cite{crack}).
In particular,
for a~given nonlinear 2nd order PDE having two independent variables,
solutions are produced in an explicit form,
whereas standard symmetry reduction only gives a 2nd order ODE
that still has to be solved to f\/ind group-invariant solutions explicitly
and in general this step can be quite dif\/f\/icult.

Moreover,
because the group foliation equations contain
all solutions of the given nonlinear PDE,
the method can yield solutions that are not invariant
under any of the point (or contact) symmetries admitted by the nonlinear PDE.

It is straightforward to extend this method to higher-order PDEs
with power nonlinearities.
As well, it should be possible to apply the same method to PDEs
having more general forms of nonlinearities
by utilizing a general separation of variables ansatz
with respect to all of the independent variables
in the group foliation equations.
In general,
the success of such an ansatz can be interpreted mathematically~\cite{Galaktionov-book}
as constructing a (partially) invariant linear subspace for
a nonlinear operator coming from the structure of the group foliation equations.

Related work using a similar group-foliation method
applied to nonlinear dif\/fusion equations appears in \cite{QuZhang}.
Group foliation equations were f\/irst used successfully
in \cite{Golovin,NutkuSheftel,SheftelWinternitz,Sheftel}
for obtaining exact solutions to nonlinear PDEs
by a dif\/ferent method that is applicable
when the group of point symmetries of a given PDE is inf\/inite-dimensional,
compared to the example of a f\/inite-dimensional symmetry group
considered both in \cite{AncoLiu,AncoAliWolf} and in the present work.

\subsection*{Acknowledgements}
S.~Anco and T.~Wolf are each supported by an NSERC research grant.
S.~Ali thanks the Mathematics Department of Brock University for support
during the period of a research visit when this paper was written.
Computations were partly performed on computers of the Sharcnet consortium (www.sharcnet.ca).
The referees and the editor are thanked for valuable comments
which have improved this paper.

\pdfbookmark[1]{References}{ref}
\LastPageEnding

\end{document}